\documentclass{ws-ijmpe}
\usepackage{amsmath,amssymb}
\usepackage{graphicx}
\usepackage{bm}
\usepackage{hyperref}
\usepackage[super,compress]{cite}

\newcommand{\sfLambda}{\mathsf{\Lambda}}
\newcommand{\sfB}{\mathsf{B}}
\newcommand{\sfI}{\mathsf{I}}
\newcommand{\sfH}{\mathsf{H}}
\newcommand{\Hvac}{\mathsf{H}_\mathrm{vac}}
\newcommand{\Hmat}{\mathsf{H}_\mathrm{mat}}
\newcommand{\Hnu}{\mathsf{H}_{\nu\nu}}
\newcommand{\thetav}{\theta_\mathrm{v}}
\newcommand{\Gf}{G_\mathrm{F}}

\newcommand{\rmi}{\mathrm{i}}
\newcommand{\rmd}{\mathrm{d}}
\newcommand{\tr}{\mathrm{tr}}
\newcommand{\diag}{\mathrm{diag}}

\renewcommand{\Im}{\mathrm{Im}}

\newcommand{\bhv}{\boldsymbol{\hat\mathbf{v}}}
\newcommand{\bfx}{\mathbf{x}}
\newcommand{\bfp}{\mathbf{p}}
\newcommand{\bnabla}{\boldsymbol{\nabla}}

\newcommand{\trho}{\tilde\varrho}

\newcommand{\caL}{\mathcal{L}}
\newcommand{\caO}{\mathcal{O}}

\begin{document}

\markboth{Huaiyu Duan}{Collective neutrino oscillations and
spontaneous symmetry breaking}
\catchline{}{}{}{}{}

\title{Collective neutrino oscillations and
  spontaneous symmetry breaking}

\author{Huaiyu Duan}
\address{Department of Physics and Astronomy, University of New
  Mexico\\
Albuquerque, NM 87131, USA\\
duan@unm.edu}


\maketitle


\begin{abstract}
Neutrino oscillations in a hot and dense astrophysical
environment such as a core-collapse supernova pose a challenging,
seven-dimensional flavor transport problem. To make the problem even more
difficult (and 
interesting), neutrinos can experience collective oscillations through
nonlinear refraction in the dense neutrino medium in this
environment. Significant 
progress has been made in the last decade towards the understanding of 
collective neutrino oscillations in various simplified neutrino gas models
with imposed symmetries and reduced dimensions. However, a
series of recent studies seem to have ``reset'' this progress by
showing that these models may not be compatible with collective neutrino
oscillations because the latter can break the symmetries spontaneously
if they are not imposed. We review
some of the key concepts
of collective neutrino oscillations by using a few simple toy models.
We also elucidate the breaking of 
spatial and directional symmetries in these models because of
collective oscillations. 
\end{abstract}


\ccode{PACS numbers: 14.60.Pq, 97.60.Bw}


\section{Introduction}

Neutrinos are abundantly produced in hot and dense astrophysical
environments such as the early universe, core-collapse supernovae (SNe)
and black-hole accretion discs in which they are also instrumental
in the dynamical, thermal and chemical evolution of these environments.
In a SN, for example, the hot proto-neutron star (PNS) at the center quickly
cools down by emitting $\sim 10^{58}$ neutrinos in all flavors within just
$\sim 10$ seconds.\cite{Woosley:2005yv} 
Outside the PNS
electron neutrinos and antineutrinos influence the supernova
dynamics and nucleosynthesis through reactions
\begin{align}
\nu_e + n &\leftrightharpoons p + e^-, &
\bar\nu_e + p &\leftrightharpoons n + e^+.
\label{eq:reactions}
\end{align}

Because the weak-interaction states $|\nu_\beta\rangle$ ($\beta=e,\mu,\tau$)
and the mass eigen states $|\nu_i\rangle$ ($i=1,2,3$)   of the
neutrino do
not coincide with each other, neutrinos can mutate from one flavor (or
weak-interaction state) into another during
propagation, which is known as the neutrino flavor
transformation or neutrino oscillations.\cite{Agashe:2014kda}
The oscillation of neutrinos between the electron flavor and other
flavors can have important physical consequences in astrophysical
environments through the reactions in Eq.~\eqref{eq:reactions}. 

Even in the coherent regime (i.e.\ without neutrino absorption,
emission and collision) 
the transport of neutrino flavors poses a very challenging
problem.  
One of the challenges is that, in an environment with a large neutrino
flux, the flavor transport can become nonlinear because of the
neutrino-neutrino coupling. In most cases the
nonlinear equations of motion (e.o.m.) which govern 
neutrino oscillations have to be solved numerically.
The computational investigations by various groups show that a dense
neutrino medium can experience collective oscillations during which
neutrinos of different momenta oscillate cooperatively.%
\cite{Kostelecky:1993dm,Pastor:2002we,Balantekin:2004ug,
Duan:2006an,Duan:2006jv,Fogli:2007bk,Duan:2007sh,Dasgupta:2009mg,
Friedland:2010sc,Mirizzi:2011tu,Galais:2011jh,Cherry:2012zw,deGouvea:2012hg,
Malkus:2014iqa}  
To obtain theoretical insights into these numerical results,
several toy models such as the bipolar model%
\cite{Kostelecky:1994dt,Duan:2005cp,Hannestad:2006nj}
have been proposed
which can be solved analytically and which bear some of the qualitative
features of the more realistic models.
Linear stability analysis
provides another useful tool to predict the physical regimes where
collective neutrino oscillations may or may not occur.%
\cite{Banerjee:2011fj}

Another challenge in solving the problem of neutrino transport is simply due
to its large dimensionality: the neutrino
transport in any real astrophysical environment even without
flavor oscillations involves 1 temporal dimension, 3 spatial
dimensions and 3 momentum dimensions.
This challenge has been largely bypassed in the previous studies
which are limited to the models with imposed
symmetries. For example, the isotropic and homogeneous
condition is usually assumed for the early universe.
However, even though such symmetries can exist in the
e.o.m.\ and may also exist in the physical systems at
some point, they may not be
preserved during neutrino oscillations.\cite{Raffelt:2013rqa}
Indeed, a series of recent work
suggest that these symmetries can be broken spontaneously by neutrino
oscillations.%
\cite{Mirizzi:2013rla,Raffelt:2013isa,Duan:2013kba,Mangano:2014zda,
Duan:2014gfa,Mirizzi:2015fva}

The goals of this paper are to review
some of the key concepts of collective neutrino oscillations through
the bipolar neutrino gas model and to explain why spatial and 
directional symmetries can be broken spontaneously by collective oscillations.
We will leave out many important topics such as
three-flavor oscillations%
\cite{Dasgupta:2007ws,Duan:2008za,Friedland:2010sc} some of which
can be found in an earlier review.\cite{Duan:2010bg}

Throughout this paper we adopt the natural units with
$\hbar=c=1$. We also assume that 
neutrinos are relativistic with speed $v\approx 1$.

\section{Collective neutrino oscillations}

\subsection{Equation of motion}
Assuming the validity of the mean-field theory,%
\cite{Friedland:2003eh,Pehlivan:2011hp}
we will use the (Wigner-transformed) neutrino (flavor) density
matrices\cite{Sigl:1992fn} 
$\rho_\bfp(t,\bfx)$ and $\bar\rho_\bfp(t,\bfx)$
to describe the flavor content of the neutrino and antineutrino of
momentum $\bfp$ at time $t$ and position $\bfx$. For simplicity we
will consider neutrino oscillations between the electron flavor and
another active flavor, say $\tau$. In the
weak-interaction basis (or flavor basis),
\begin{align}
\rho_\bfp &= \begin{bmatrix}
\rho^{ee}_\bfp &  \rho^{e\tau}_\bfp \\
(\rho^{e\tau}_\bfp)^* &  \rho^{\tau\tau}_\bfp 
\end{bmatrix}, &
\bar\rho_\bfp &= \begin{bmatrix}
\bar\rho^{ee}_\bfp &  \bar\rho^{e\tau}_\bfp \\
(\bar\rho^{e\tau}_\bfp)^* &  \bar\rho^{\tau\tau}_\bfp 
\end{bmatrix}. 
\label{eq:rho}
\end{align} 
The diagonal elements of the density matrices in the above equation
are proportional to the number densities of the neutrino or antineutrino in the
corresponding flavors and momentum state, and
\begin{align}
n_{\nu_\beta}(t,\bfx) &= \int\frac{\rmd^3
                    p}{(2\pi)^3}\rho^{\beta\beta}_\bfp(t,\bfx), &
n_{\bar\nu_\beta}(t,\bfx) &= \int\frac{\rmd^3
                    p}{(2\pi)^3}\bar\rho^{\beta\beta}_\bfp(t,\bfx)
\end{align}
are the total number densities of the neutrinos and antineutrinos in
flavor $\beta$ ($\beta=e,\tau$), respectively. The off-diagonal elements of the
density matrices contain the information of flavor mixing. In the rest
of the paper we will use the weak-interaction basis exclusively unless
otherwise stated.

In the coherent regime a neutrino can
change its flavor through refraction when it passes
through a dense medium. 
In the lowest-order and without gravitational redshift,
the e.o.m.\ that govern the neutrino flavor
transformation are%
\cite{Sigl:1992fn,Strack:2005ux,Cardall:2007zw}
\begin{subequations} 
\label{eq:eom-old}
\begin{align}
(\partial_t + \bhv\cdot\bnabla)\rho_\bfp &= 
-\rmi [\Hvac+\Hmat+\Hnu,\, \rho_\bfp], \\
(\partial_t + \bhv\cdot\bnabla)\bar\rho_\bfp &= 
-\rmi [-\Hvac+\Hmat+\Hnu,\, \bar\rho_\bfp], 
\end{align}
\end{subequations}
where $\bhv=\bfp/E$ is the propagation velocity of the neutrino with
$E=|\bfp|$ being the energy of the neutrino.
The vacuum Hamiltonian is
\begin{align}
\Hvac = \frac{\Delta m^2}{4 E}
\begin{bmatrix}
-\cos2\thetav & \sin2\thetav \\
\sin2\thetav & \cos2\thetav 
\end{bmatrix},
\end{align}
where $\Delta m^2$ is the neutrino mass-squared difference, and
$\thetav$ is the vacuum mixing angle within range
$(0,\pi/4]$. 
Here $\Delta m^2 > 0$ and $\Delta m^2<0$ correspond to
the normal neutrino mass hierarchy (NH) and the inverted hierarchy
(IH), respectively.
The matter potential in Eq.~\eqref{eq:eom-old} arises from the
coherent forward scattering of the neutrino by the charged leptons in the
medium through the charged-current weak
interaction.\cite{Wolfenstein:1977ue}
When the
densities of $\mu$ and $\tau$ leptons are negligible, the matter
potential can be written as
\begin{align}
\Hmat = \sqrt2\Gf \begin{bmatrix}
n_e & 0 \\ 0 & 0 \end{bmatrix}
= \frac{\lambda}{2}(\sfI + \sigma_3),
\end{align}
where $\Gf$ is the Fermi constant, $n_e$ is the net electron
number density, $\lambda=\sqrt2\Gf n_e$, $\sfI$ is the identity
matrix, and $\sigma_3$ is the third Pauli matrix. 
The neutrino(-neutrino coupling) potential in Eq.~\eqref{eq:eom-old}
stems from the coherent forward 
scattering of the neutrino in question by ambient neutrinos through
the neutral-current interaction%
\cite{Fuller:1987aa,Notzold:1987ik,Pantaleone:1992xh}, and it can be
written as
\begin{align}
\Hnu = \sqrt2\Gf \int\!\frac{\rmd^3 p'}{(2\pi)^3}
(1-\bhv\cdot\bhv')(\rho_{\bfp'} - \bar\rho_{\bfp'}),
\label{eq:Hnu}
\end{align}
where the physical quantities with primes are for the ambient
neutrinos.

We will be interested in the regimes with large matter densities where
the (effective) neutrino mixing angle in matter is small and
\begin{align}
\sfB_\omega = \Hvac + \Hmat \xrightarrow{\sqrt2 \Gf n_e \gg |\Delta
  m^2/2 E|}
(\lambda - \eta\omega)
  \frac{\sigma_3}{2},
\label{eq:B}
\end{align}
where $\eta=+1$ and $-1$ for NH and IH, respectively, and 
\begin{align}
\omega(E) = \frac{|\Delta m^2|}{2 E}\cos2\thetav.
\end{align}
We have ignored the trace terms here which have no impact on neutrino
oscillations. 
For antineutrino,
\begin{align}
\bar\sfB_{\omega} = -\Hvac + \Hmat \longrightarrow (\lambda + \eta\omega)
  \frac{\sigma_3}{2} = \sfB_{-\omega}.
\label{eq:Bbar}
\end{align}
Therefore, for the purpose of neutrino oscillations one can treat
an antineutrino of oscillation frequency $\omega$ and energy $E$ as a
neutrino with oscillation frequency $-\omega$ and energy $-E$. We will
adopt this convention throughout this paper.

In the absence of neutrino absorption, emission and collision,
it is useful to define (reduced) neutrino flavor (density) matrices
\begin{align}
\varrho_{\omega>0,\bhv}&\propto\rho_\bfp, &
\varrho_{\omega<0,\bhv} &\propto \bar\rho_\bfp
\end{align}
with normalization condition
\begin{align}
\tr\varrho_{\omega,\bhv} = 1.
\end{align}
The diagonal elements of $\varrho_{\omega,\bhv}$ give the
probabilities for the corresponding neutrino ($\omega>0$) or
antineutrino ($\omega<0$) in the corresponding weak-interaction states.
Neutrino flavor matrices $\varrho_{\omega,\bhv}$ obey the e.o.m.
\begin{align}
(\partial_t + \bhv\cdot\bnabla)\varrho_{\omega,\bhv}
=-\rmi[\sfB_\omega+\Hnu,\, \varrho_{\omega,\bhv}].
\label{eq:eom}
\end{align}

\subsection{Bipolar model%
\label{sec:bipolar}}

The study of collective neutrino oscillations begins with the bipolar model.%
\cite{Kostelecky:1994dt,Duan:2005cp,Hannestad:2006nj}
This model describes a homogeneous and isotropic system with
$\nu_e$ and $\bar\nu_e$ populations of single energy $E_0$ at time 
$t=0$. The e.o.m.\ of neutrino oscillations is
\begin{align}
\rmi \partial_t\varrho_{\omega}
=\left[(\lambda-\eta\omega)\frac{\sigma_3}{2}+\Hnu,\, 
\varrho_{\omega}\right]
\label{eq:eom-bipolar}
\end{align}
with $\omega=\pm\omega_0$, and the neutrino potential is
\begin{align}
\Hnu = \mu (\varrho_{\omega_0} - \alpha\varrho_{-\omega_0}),
\end{align}
where
\begin{align}
  \alpha=\frac{n_{\bar\nu}}{n_{\nu}}
  \label{eq:alpha}
\end{align}
is the ratio of the density of 
the antineutrino to that of the neutrino, and
\begin{align}
  \mu = \sqrt 2 \Gf n_{\nu}
\end{align}
is a measure of the strength of the neutrino-neutrino coupling.

If the neutrino density is negligible, 
Eq.~\eqref{eq:eom-bipolar} has solution
\begin{align}
\varrho_{\pm\omega_0}(t) &\approx 
\exp\left[-\rmi(\lambda\mp\eta\omega_0)t\frac{\sigma_3}{2}\right]
\varrho_{\pm\omega_0}(0) 
\exp\left[\rmi(\lambda\mp\eta\omega_0)t\frac{\sigma_3}{2}\right].
\end{align}
Because 
\begin{align}
  \varrho_{\pm\omega_0}(0)
=\begin{bmatrix}
1 & 0 \\ 0 & 0 \end{bmatrix}
\end{align}
are diagonal and commute with $\sigma_3$, one has
\begin{align}
  \varrho_{\pm\omega}(t)\approx
\begin{bmatrix}
1 & 0 \\ 0 & 0 \end{bmatrix}.
\end{align}
In other words, neutrino oscillations are suppressed in the presence
of dense matter if there is no ambient neutrino.

However, the presence of dense matter does not necessarily suppress collective
oscillations in the dense neutrino medium. This can be seen by
performing a ``corotating-frame'' transformation to the neutrino
flavor matrix:\cite{Duan:2005cp}
\begin{align}
\trho_\omega(t) = e^{\rmi\lambda t\sigma_3/2}\rho_\omega(t)
e^{-\rmi\lambda t\sigma_3/2},
\end{align}
which removes the time evolution
due to matter. Under this transformation Eq.~\eqref{eq:eom-bipolar} becomes
\begin{align}
\rmi\partial_t\trho_\omega(t) = \left[
-\eta\omega\frac{\sigma_3}{2}+\tilde\sfH_{\nu\nu},\,\trho_\omega\right],
\label{eq:eom-tbipolar}
\end{align}
where $\tilde\sfH_{\nu\nu} = \mu (\trho_{\omega_0} -
\alpha\trho_{-\omega_0})$. 
Eqs.~\eqref{eq:eom-bipolar} and 
\eqref{eq:eom-tbipolar} are equivalent, and 
$\varrho$ and $\trho$ have the same diagonal elements. Therefore,
if there exists collective flavor transformation in the
bipolar model, a uniform dense matter will not suppress it.%
\footnote{However, collective neutrino oscillations can be suppressed
  by a very large matter density in SNe as shown in
  Ref.~\citen{EstebanPretel:2008ni}.}
In the rest of the paper we will always work in the corotating frame
by taking $\lambda=0$, but we
will drop the tilde for simplicity.

To see that a bipolar system can experience collective flavor
transformation we note that Eq.~\eqref{eq:eom-tbipolar} is invariant under
global phase transformation\cite{Duan:2008fd} 
\begin{align}
\varrho_\omega(t) \longrightarrow e^{-\rmi\phi\sigma_3/2}
  \varrho_\omega(t) e^{\rmi\phi\sigma_3/2},
\label{eq:sym}
\end{align}
where $\phi$ is a constant. This symmetry leads to the conservation of
the ``flavor lepton number (density)''\cite{Raffelt:2007cb}
\begin{align}
\caL
= n_\nu \tr[(\varrho_{\omega_0}-\alpha\varrho_{-\omega_0})\sigma_3]
= (n_{\nu_e}-n_{\nu_\tau}) - (n_{\bar\nu_e}-n_{\bar\nu_\tau}).
\label{eq:L}
\end{align}
The phase symmetry in Eq.~\eqref{eq:sym} also implies the existence
of a collective precession solution:
\begin{align}
\varrho_\omega(t) = 
e^{-\rmi\Omega t\sigma_3/2}
\varrho_{\omega}(0) e^{\rmi\Omega t\sigma_3/2},
\label{eq:sol-coll}
\end{align}
where $\Omega$ is a common oscillation frequency independent
of the oscillation frequency $\omega$ of the individual neutrino.

The bipolar model can be solved analytically.\cite{Kostelecky:1994dt}
It has also been shown that the bipolar model
is equivalent to a gyroscopic pendulum in flavor
space,\cite{Hannestad:2006nj} and the collective precession solution in
Eq.~\eqref{eq:sol-coll} corresponds to the precession motion of this
pendulum. However, this collective precession solution can not be 
achieved exactly for arbitrary initial conditions 
just as a gyroscopic pendulum usually experiences both nutation and
precession. 

\begin{figure*}[ht]
  \begin{center}
    $\begin{array}{cc}
      \includegraphics*[width=0.5\textwidth]{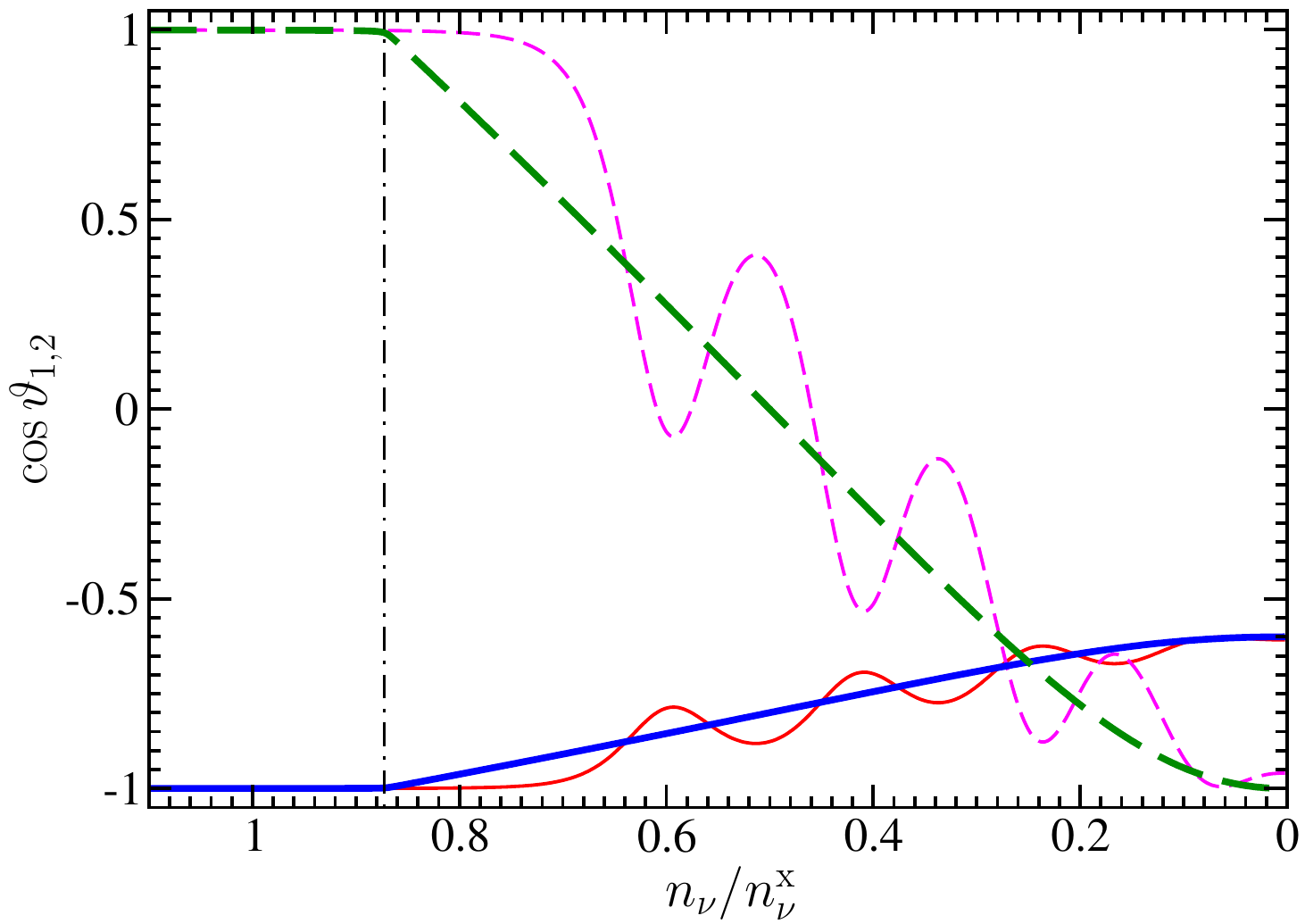} &
      \includegraphics*[width=0.5\textwidth]{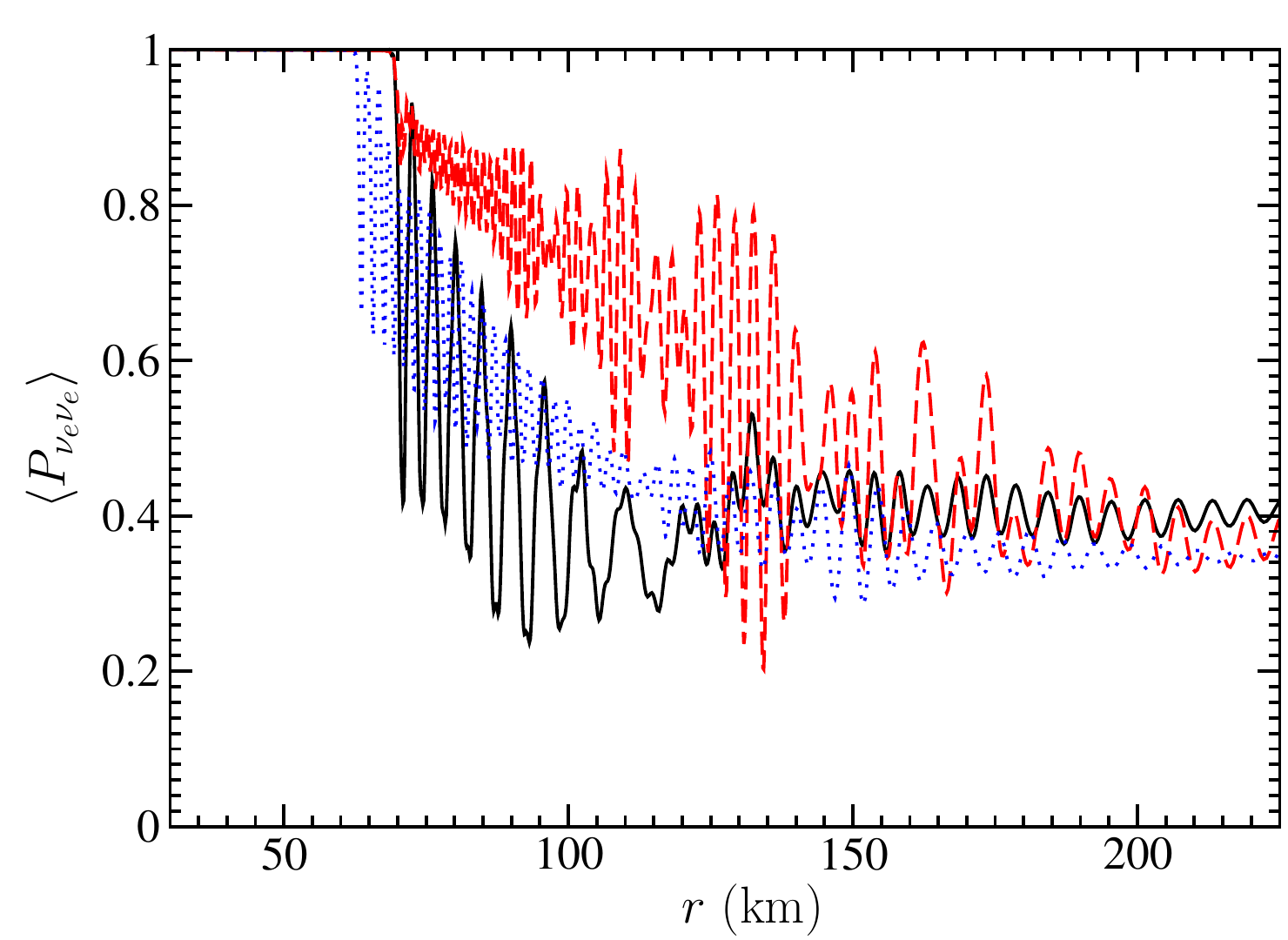}
      \end{array}$
  \end{center}
  \caption{Left panel: The collective precession solution of a bipolar model
    with time-varying neutrino density $n_\nu$
    (thick lines) and the numerical solution for this model (thin
    lines). The dashed and solid lines represent
    $\cos\vartheta_1 = \tr(\varrho_{\omega_0}\sigma_3)$ and 
    $\cos\vartheta_2 = -\tr(\varrho_{-\omega_0}\sigma_3)$, respectively. 
    Right panel: The energy-averaged neutrino survival probabilities
    for three representative neutrino trajectories as functions of
    radius $r$ in a spherical supernova model.
    The figures are adapted from Fig.~3 of Ref.~\citen{Duan:2007mv} and Fig.~2
    of Ref.~\citen{Duan:2006jv}, respectively.}
  \label{fig:bipolar-time}
\end{figure*}

The bipolar model is, of course, very different from the neutrino
medium in a realistic astrophysical environment. However, it does
provide explanations for some of the important results of
collective neutrino oscillations in more sophisticated models. 
For example, one can show that in the collective
precession solution neutrinos will remain in the weak-interaction
states if the neutrino density is larger than a critical value%
\cite{Hannestad:2006nj} [see the discussion in
Sec.~\ref{sec:instability} and around Eq.~\eqref{eq:mu-range}].
This corresponds to the ``sleeping-top'' regime of a gyroscopic
pendulum where the fast spinning top or pendulum defies gravity
and spins without wobbling. When the neutrino number density decreases below the
critical value, the flavor pendulum begins to wobble about the precession
solution. Similar phenomenon was indeed observed in numerical simulations of
neutrino oscillations in SNe (see
Fig.~\ref{fig:bipolar-time}). 

\subsection{Spectral swap/split}

The collective precession solution in Eq.~\eqref{eq:sol-coll} also exists
in a homogeneous and isotropic neutrino gas with continuous neutrino energy
spectra.\cite{Raffelt:2007cb}
We will again assume that the neutrinos and antineutrinos are in the
weak-interaction states at time $t=0$. Because Eq.~\eqref{eq:eom-old}
is invariant under $\rho\rightarrow-\rho$ (without changing
$\sfH_{\nu\nu}$), we can use
\begin{align}
\varrho_\omega(t=0) = \begin{bmatrix}
1 & 0 \\ 0 & 0 \end{bmatrix}
\end{align}
as the initial condition for all neutrinos and antineutrinos by
replacing the neutrino energy spectra with a single effective spectrum
\begin{align}
g(\omega) \propto \left|\frac{\rmd E}{\rmd\omega}\right|\times
\left\{\begin{array}{ll}
\tr(\rho_E\sigma_3)_{t=0} & \text{ if } \omega > 0, \\
-\tr(\bar\rho_E\sigma_3)_{t=0} & \text{ if } \omega < 0
\end{array}\right.
\end{align}
with normalization%
\footnote{Other equivalent normalization conditions also exist in
  the literature. The normalization condition in Eq.~\eqref{eq:g-norm} is
  used in anticipation that the $\nu_e$ flux is the largest among all
  flavors in a SN.}
\begin{align}
  \int_0^\infty g(\omega)\,\rmd\omega = 1.
  \label{eq:g-norm}
\end{align}
The e.o.m.\ \eqref{eq:eom-tbipolar} also applies to the
neutrino gas with continuous neutrino
energy spectra except that the neutrino potential is now
\begin{align}
\Hnu = \mu \int_{-\infty}^\infty g(\omega) \varrho_\omega\,\rmd\omega,
\end{align}
where
\begin{align}
  \mu = \sqrt2\Gf (n_{\nu_e} - n_{\nu_\tau})_{t=0}.
  \label{eq:mu}
\end{align}
The definition of $\alpha$ in Eq.~\eqref{eq:alpha} can be
generalized to
\begin{align}
  \alpha = -\frac{\int_{-\infty}^0 g(\omega)\,\rmd\omega}%
         {\int_0^\infty g(\omega)\,\rmd\omega}.
\end{align}
Obviously, the bipolar model is a special case of the
continuous-spectrum model with
\begin{align}
g(\omega) = \delta(\omega-\omega_0) - \alpha \delta(\omega+\omega_0).
\label{eq:g-bipolar}
\end{align}

\begin{figure*}[ht]
  \begin{center}
    $\begin{array}{cc}
      \includegraphics*[width=0.45\textwidth]{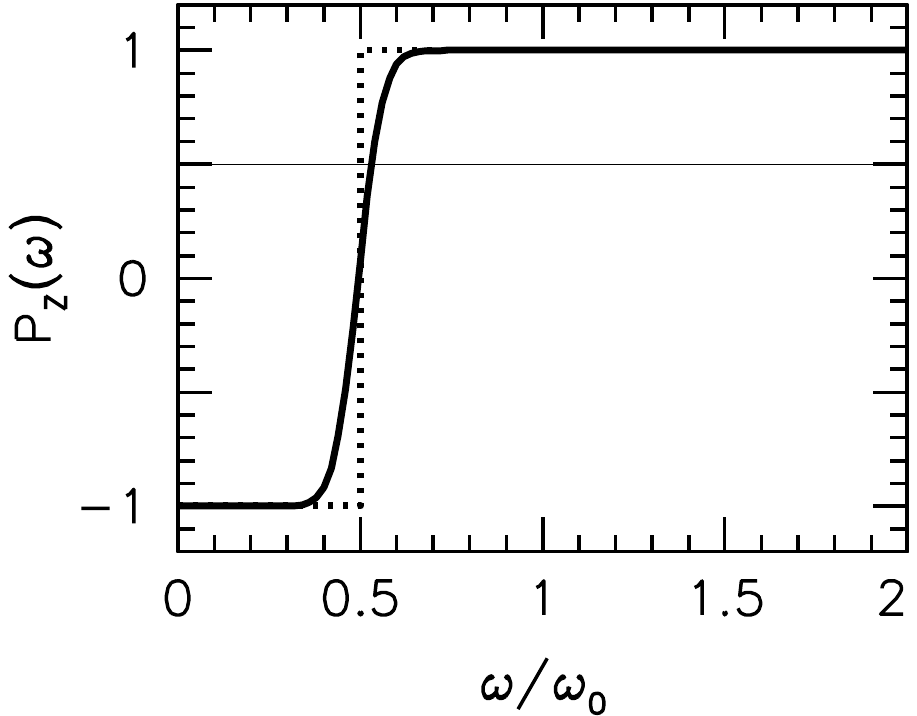} &
      \includegraphics*[width=0.55\textwidth]{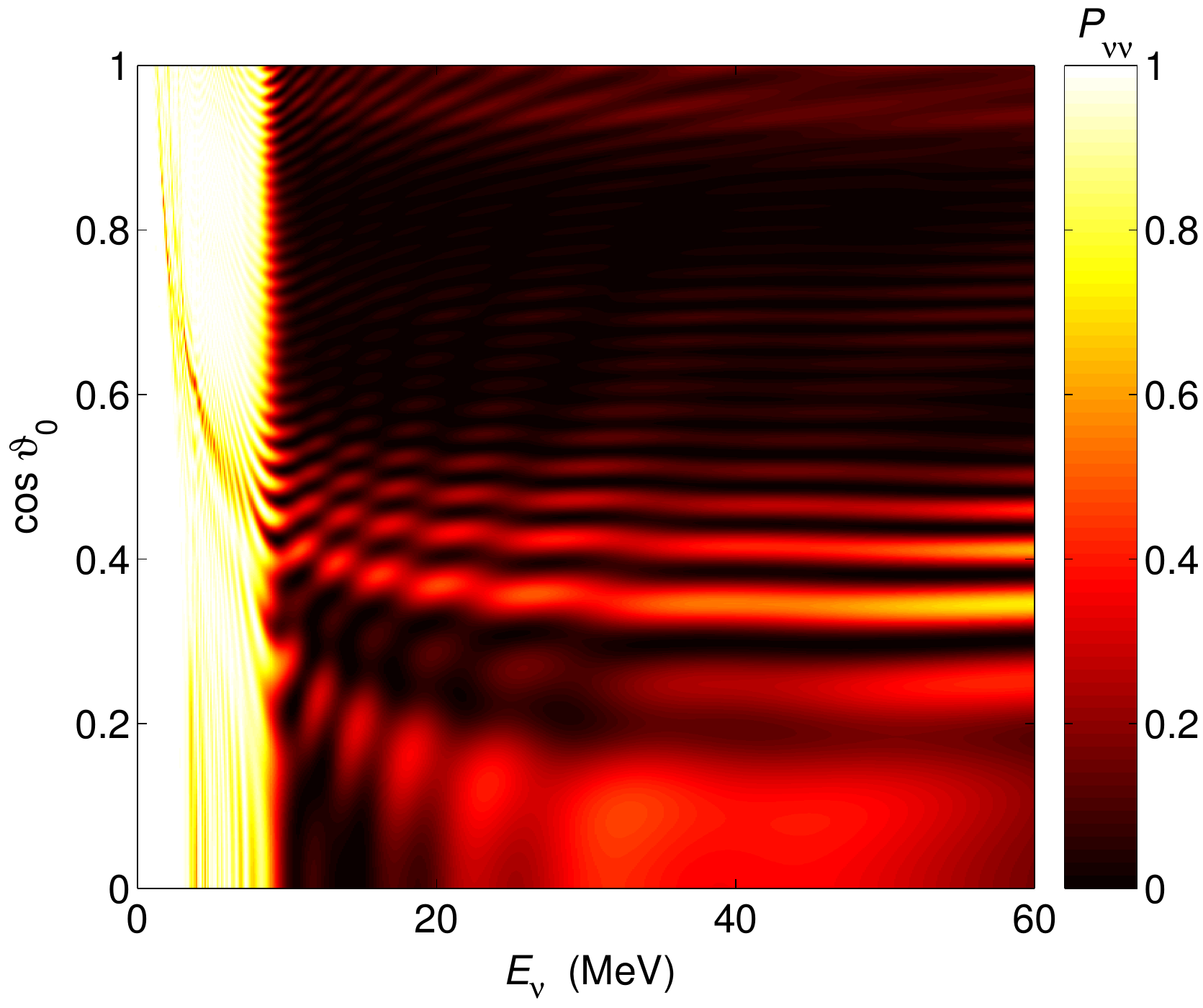}
      \end{array}$
  \end{center}
  \caption{Left panel: The split neutrino spectrum (dotted line: fully
    adiabatic; thick solid line: numerical) produced from a box-like
    initial spectrum (thin solid line)
    in a homogeneous and isotropic neutrino gas
    by collective neutrino oscillations,
    where $\omega=|\Delta m^2|/2 E_\nu$ and $P_z(\omega) =
    \tr(\varrho_\omega)$. Right panel: The final neutrino
    survival 
    probability as a function of neutrino energy $E_\nu$ and neutrino
    emission angle $\vartheta_0$ on the surface of the proto-neutron
    star in a numerical 
    calculation employing a spherical supernova model.
    The figures are adapted from Fig.~1 of Ref.~\citen{Raffelt:2007cb}
    and Fig.~3 of Ref.~\citen{Duan:2006jv}, respectively.
    \label{fig:split}}
\end{figure*}

If the neutrino gas is in the collective precession mode initially and
expands adiabatically afterwards, it should remain in the collective mode.
However, according to Eq.~\eqref{eq:eom-tbipolar}, when the neutrino
density becomes negligible  (at time $t\geq
t_1$), 
neutrinos of different energies should oscillate
independently with their own oscillation frequencies:
\begin{align}
\varrho_\omega(t) = e^{\rmi\eta\omega(t-t_1)\sigma_3/2} \varrho_\omega(t_1)
e^{-\rmi\eta\omega(t-t_1)\sigma_3/2}.
\end{align}
This apparent contradiction is resolved if $\varrho_\omega(t_1)$ is
diagonal. Being diagonal does not necessarily mean that
$\varrho_\omega(t_1)=\varrho_\omega(0)=\diag[1,0]$. For example,
if the neutrino gas has a positive flavor lepton number $\caL$ [see
  Eq.~\eqref{eq:L}] and
a bipolar-like spectrum [i.e.\ 
$g(\omega)$ is positive (negative) if $\omega>0$ ($\omega<0$)] and if the 
neutrino mass hierarchy is inverted,
the final spectrum can be split at the critical energy
\begin{align}
  E_\text{C} = \frac{|\Delta m^2|}{2\Omega_0}
\end{align}
such that
\begin{align}
  \varrho_\omega(t_1) = \left\{\begin{array}{ll}
  \diag[1,0] & \text{ if } \omega > \Omega_0,\\
  \diag[0,1] & \text{ if } \omega < \Omega_0,
  \end{array}\right.
\end{align}
where $\Omega_0$ is the collective oscillation frequency at
$\mu=0$\cite{Duan:2006an} and can be determined by the conservation of
$\caL$\cite{Raffelt:2007cb}. The above result implies that $\nu_e$ 
and $\nu_\tau$ have swapped their energy spectra at energies $E > E_\text{C}$.
This spectral swap/split phenomenon is commonly
observed in the numerical simulations of neutrino oscillations in SNe (see
Fig.~\ref{fig:split}).
If the neutrino gas has a more complicated spectra with multiple
spectral crossings where $g(\omega)=0$, the final spectra can exhibit multiple
  spectral splits.\cite{Dasgupta:2009mg}.

\subsection{Flavor instability%
\label{sec:instability}}

Although the bipolar model can be solved analytically and provide some
qualitative understandings of 
collective oscillations in neutrino media, numerical
computations are usually required to obtain quantitative results of
neutrino oscillations in more sophisticated models.
However, if the neutrinos and antineutrinos are initially in
the weak-interaction states, linear stability analysis can be utilized
to predict the physical regimes where collective oscillations can
occur without performing large-scale 
numerical simulations.
This is especially useful if a
large parameter space needs to be surveyed
because of the uncertainty in, e.g., the neutrino fluxes in SN simulations. 
The linear stability analysis method was explained thoroughly in
Ref.~\citen{Banerjee:2011fj}. Here we will use the bipolar model to
demonstrate the essence of this method.

We assume that no significant flavor oscillations have
occurred at time $t$ such that
\begin{align}
\varrho_{\omega_0} &\approx \begin{bmatrix}
1 &  \epsilon \\  \epsilon^* & 0 \end{bmatrix}, &
\varrho_{-\omega_0} &\approx \begin{bmatrix}
1 & \bar \epsilon \\ \bar \epsilon^* & 0 \end{bmatrix}, 
\label{eq:eps}
\end{align}
where $| \epsilon|\sim|\bar \epsilon|\ll 1$. Keeping only the terms
up to $\caO(\epsilon)$ we can rewrite
Eq.~\eqref{eq:eom-tbipolar} in terms of 
$\epsilon$ and $\bar\epsilon$ as
\begin{align}
\rmi\partial_t\begin{bmatrix} \epsilon \\ \bar \epsilon \end{bmatrix}
&\approx
\begin{bmatrix}
-\eta\omega-\mu\alpha & \alpha\mu \\
-\mu & \eta\omega +\mu \end{bmatrix}
\begin{bmatrix} \epsilon \\ \bar \epsilon \end{bmatrix}
=\sfLambda \cdot\begin{bmatrix}\epsilon \\ \bar\epsilon \end{bmatrix}
  .
\label{eq:bipolar-lin}
\end{align}
Note that, in the absence of the neutrino medium (i.e.\ $\mu=0$),
$\epsilon$ and $\bar\epsilon$ are decoupled and oscillate with
frequencies $-\eta\omega$ and $\eta\omega$, respectively.  In the
neutrino medium, however, $\epsilon$ and $\bar\epsilon$ are coupled,
and each eigenvalue $\Omega$ of matrix $\sfLambda$ corresponds to a
collective oscillation mode:
\begin{align}
\epsilon(t) &\approx Q e^{-\rmi\Omega t}, &
\bar\epsilon(t) &\approx \bar{Q} e^{-\rmi\Omega t},
\label{eq:sol-eps}
\end{align}
where $[Q,\bar Q]^T$ is the corresponding eigenvector of $\sfLambda$.

If $\Omega$ is real, $\epsilon$ and $\bar\epsilon$ will oscillate and
their amplitudes will remain small.
If $\Omega$ is complex with
\begin{align}
\kappa = \Im(\Omega) > 0,
\end{align}
the bipolar model has a flavor instability: the
amplitudes of $\epsilon$ and $\bar\epsilon$ will grow 
exponentially which can lead to (large-magnitude) collective
oscillations. Because  
$\sfLambda$ is a real matrix, when $\Omega$ is complex, $\Omega^*$ is
also an eigenvalue of $\sfLambda$. In this case, the solution
with positive $\kappa$ will dominate over time. Therefore, for the
purpose of flavor stability analysis, it is sufficient
to find out the regimes where $\sfLambda$ has complex eigenvalues.

It is straightforward to show that $\sfLambda$ has complex eigenvalues
only when
\begin{align}
  (1-\alpha)^2\mu^2 + 4\omega_0^2 + 4(1+\alpha)\mu\eta\omega_0 < 0
\end{align}
or
\begin{align}
\frac{2\omega_0}{(1+\sqrt\alpha)^2 }
< -\eta \mu <
\frac{2\omega_0}{(1-\sqrt\alpha)^2}.
\label{eq:mu-range}
\end{align}
The bipolar model (with $\nu_e$ and $\bar\nu_e$ initially)
can have flavor instability only in IH.
The upper limit of $\mu$ where collective oscillations can occur
indeed agrees with the result plotted in Fig.~\ref{fig:bipolar-time}(a).
It is interesting to note that, if the bipolar model has
$\nu_\tau$ and $\bar\nu_\tau$ initially instead of $\nu_e$ and $\bar\nu_e$, it
will have flavor instability only in NH because $\mu<0$ [see Eq.~\eqref{eq:mu}].

Similarly, for the continuous-spectrum model, we assume
\begin{align}
\varrho_\omega \approx \begin{bmatrix}
1 &  \epsilon_\omega \\  \epsilon_\omega^* & 0 
\end{bmatrix},
\label{eq:rho0}
\end{align}
the e.o.m.\ \eqref{eq:eom-tbipolar} can be written as
\begin{align}
\rmi\partial_t \epsilon_\omega \approx
[-\eta\omega + (1-\alpha)\mu] \epsilon_\omega 
-\mu\int_{-\infty}^\infty g(\omega') \epsilon_{\omega'}\,\rmd\omega'.
\label{eq:eps-cont}
\end{align}
Using the ansatz 
\begin{align}
 \epsilon_\omega(t) =   Q_\omega e^{-\rmi\Omega t}
\end{align}
one obtains
\begin{align}
  [\Omega -(1-\alpha)\mu+\eta\omega] Q_\omega = \text{const.}
  \qquad\text{or}\qquad
Q_\omega \propto \frac{1}{\Omega-(1-\alpha)\mu +\eta\omega}.
\end{align}
In order for the above solution be consistent with
Eq.~\eqref{eq:eps-cont}, $\Omega$ must satisfy the
consistency condition 
\begin{align}
-\frac{1}{\mu} =
\int_{-\infty}^\infty\frac{g(\omega)}{\Omega-(1-\alpha)\mu+\eta\omega}
\,\rmd\omega .
\end{align}
Because $\widetilde\Omega = \Omega - (1-\alpha)\mu$
and $\Omega$ have the same imaginary part, it is sufficient to solve
\begin{align}
-\frac{1}{\mu} =
  \int_{-\infty}^\infty\frac{g(\omega)}{\widetilde\Omega+\eta\omega}\,\rmd\omega 
\label{eq:consistency}
\end{align}
for $\widetilde\Omega$ for the purpose of flavor stability analysis.
A complex $\widetilde\Omega$ implies the existence of a flavor
instability. 

Flavor stability analysis can also be applied to more complicated
models. Although a powerful tool, the linear stability
analysis method depends on the validity of assumption
\eqref{eq:rho0} and is limited to the linear regime. When there is a
flavor instability, numerical simulations are still needed to follow
neutrino oscillations in the nonlinear regime.

\section{Spontaneous symmetry breaking}

The e.o.m.\ \eqref{eq:eom} involves seven dimensions and has not
been solved in its complete form. Instead, various symmetries have been
assumed to reduce the dimensionality of the problem so that a
numerical or analytically solution can be
found. Three classes of symmetries are commonly employed: the time translation
symmetry that makes the problem time-independent, spatial symmetries
that reduce the spatial dimensions, and directional symmetries that
reduce the momentum dimensions. For example, a commonly used model for
the early universe assumes complete (spatial) homogeneity and
(directional) isotropy\cite{Kostelecky:1993dm,Abazajian:2002qx}. 
These assumptions reduce the neutrino
transport in the early universe to a two-dimensional problem. Another
example is the SN
neutrino Bulb model\cite{Duan:2006an} which assumes the time
translation symmetry, the (spatial) spherical symmetry about the center
of the SN, and the (directional) axial symmetry about the radial
direction. These assumptions reduce the neutrino transport in the SN
to a three-dimension problem.

However, the full e.o.m.\ \eqref{eq:eom} allows solutions
that may or may not have these symmetries. If  a symmetry-breaking
solution becomes unstable, collective oscillations can break the
corresponding symmetry spontaneously even if such an (approximate)
symmetry exists in the neutrino medium
initially\cite{Raffelt:2013rqa}. We will first use the two-beam model to
illustrate the basic idea of spontaneous symmetry breaking by
collective oscillations.
We will then look at the breaking of directional and spatial symmetries
separately.

\subsection{Two-beam model%
\label{sec:two-beam}}

We consider a model in which all neutrinos are emitted from the $x$
axis and propagate only in the $x$-$z$ plane. We assume that neutrinos
are emitted in only two directions:
\begin{align}
\bhv_\zeta = [u_\zeta,0,v_z]
\qquad(\zeta=L,R),
\end{align}
where $0<v_z<1$ and $u_R=-u_L=\sqrt{1-v_z^2}$. We also assume that
every  point on the $x$ axis emits  in each direction neutrinos and
antineutrinos of single energy $E_0$ with intensities $j_{\nu}$ and 
$j_{\bar\nu}=\alpha j_{\nu}$, respectively. We further impose
the time translation symmetry and the translation symmetry along the $x$
axis. With these conditions Eq.~\eqref{eq:eom} reduces to
\begin{align}
\rmi \bhv_\zeta\cdot\bnabla\varrho_{\omega,\zeta}
=\left[-\eta\omega\frac{\sigma_3}{2} 
+\Hnu
,\, \varrho_{\omega,\zeta}\right].
\label{eq:two-beam}
\end{align}
In the above equation the neutrino potential is
\begin{subequations}
\begin{align}
\Hnu &= \sqrt2\Gf j_\nu
\sum_{\zeta'} (1-\bhv_\zeta\cdot\bhv_{\zeta'})
\int_{-\infty}^\infty g(\omega') \varrho_{\omega',\zeta'}\,\rmd\omega'\\
&=\mu(\varrho_{\omega_0,\tilde\zeta} -\alpha\varrho_{-\omega_0,\tilde\zeta}),
\end{align}
\end{subequations}
where
\begin{align}
\mu = \sqrt2(1-\bhv_L\cdot\bhv_R)\Gf j_\nu,
\end{align}
and $\tilde\zeta=R,L$ are the opposites of $\zeta$.

The two beam model has been solved analytically.\cite{Raffelt:2013isa}
Here we use the linear stability analysis method to demonstrate that the
left-right symmetry ($L\leftrightarrow R$) can be broken
spontaneously.\cite{Duan:2013kba} 
We again assume that the neutrinos and antineutrinos are in the
(almost pure) electron flavor:
\begin{align}
\varrho_{\omega_0,\zeta} &\approx \begin{bmatrix}
1 &  \epsilon_\zeta \\  \epsilon^*_\zeta & 0 \end{bmatrix}, &
\varrho_{-\omega_0,\zeta} &\approx \begin{bmatrix}
  1 & \bar \epsilon_\zeta \\ \bar \epsilon^*_\zeta & 0 \end{bmatrix}.
\label{eq:rho0-two-beam}
\end{align}
We then define
\begin{align}
  \epsilon^\pm &= \frac{\epsilon_L\pm\epsilon_R}{\sqrt2}, &
  \bar\epsilon^\pm &= \frac{\bar\epsilon_L\pm\bar\epsilon_R}{\sqrt2}.
  \label{eq:eps-pm}
\end{align}
Keeping only the terms
up to $\caO(\epsilon)$  we can rewrite Eq.~\eqref{eq:two-beam} as
\begin{align}
\rmi \partial_z\begin{bmatrix}
\epsilon^+ \\ \bar\epsilon^+ \\ \epsilon^- \\ \bar\epsilon^-
\end{bmatrix}
\approx
\begin{bmatrix}
\sfLambda_+ & \\
& \sfLambda_-
\end{bmatrix}\cdot
\begin{bmatrix}
\epsilon^+ \\ \bar\epsilon^+ \\ \epsilon^- \\ \bar\epsilon^-
\end{bmatrix},
\end{align}
\label{eq:two-beam-lin}
where
\begin{subequations}
\begin{align}
\sfLambda_+ &= v_z^{-1} 
\begin{bmatrix}
-\eta\omega_0 -\alpha\mu & \alpha\mu \\
-\mu & \eta\omega_0 + \mu 
\end{bmatrix},
\\
\sfLambda_- &= v_z^{-1} 
\begin{bmatrix}
-\eta\omega_0 + (2-\alpha)\mu & -\alpha\mu \\
\mu & \eta\omega_0 + (1-2\alpha)\mu
\end{bmatrix}.
\end{align}
\end{subequations}
Therefore,
the plus modes $[\epsilon^+,\bar\epsilon^+]$ and the minus modes
$[\epsilon^-,\bar\epsilon^-]$ evolve independently in the linear
regime.

The two-beam model is a 3-dimensional model with 1 spatial
dimension ($z$) and 
2 momentum dimensions ($\omega$, $\zeta$).
Its e.o.m.\ \eqref{eq:two-beam} is invariant under the
simultaneous interchange of the left and right neutrino beams.
When this left-right symmetry
is artificially imposed, Eq.~\eqref{eq:two-beam} reduces to the
e.o.m.\ of the 2-dimensional bipolar model
[Eq.~\eqref{eq:eom-tbipolar}]. The bipolar model
permits only the symmetry-preserving solutions which can be
unstable only in IH (see Sec.~\ref{sec:instability}). However,
because
\begin{align}
  \sfLambda_-= -\sfLambda_+|_{\eta\rightarrow-\eta} + 
2(1-\alpha)\mu \sfI,
\end{align}
the minus modes in the two-beam model are unstable in the same range
of $\mu$ in NH as the plus modes in IH. In the regime where the
minus modes become unstable, the left-right symmetry in the two-beam
model is broken spontaneously, and the neutrino oscillations in the
two-beam model and the bipolar model are qualitatively different.

\subsection{Breaking of the directional symmetry}

The breaking of the left-right symmetry in the two-beam model is an
example of how the directional symmetry can be broken during neutrino
oscillations. The origin of this symmetry breaking lies in the current-current
nature of the weak coupling. This is best illustrated using the
bipolar model in Sec.~\ref{sec:bipolar} but without imposing the
isotropic condition.\cite{Duan:2013kba} 
The e.o.m.\ of this model is
\begin{align}
\rmi \partial_t\varrho_{\omega,\bhv}
=\left[-\eta\omega\frac{\sigma_3}{2}+\Hnu,\, 
\varrho_{\omega,\bhv}\right],
\label{eq:eom-gas}
\end{align}
where 
\begin{align}
  \Hnu = \frac{\sqrt2}{4\pi}\Gf n_\nu \int\,\rmd\Gamma_{\bhv'}(1-\bhv\cdot\bhv')
  (\varrho_{\omega_0,\bhv'}-\alpha\varrho_{-\omega_0,\bhv'})
\end{align}
with $\rmd\Gamma_{\bhv'}$ being the differential solid angle around
direction $\bhv'$. 
We will again assume that
\begin{align}
\varrho_{\omega_0,\bhv} &\approx\begin{bmatrix}
1 & \epsilon_{\bhv} \\
\epsilon_{\bhv}^* & 0 \end{bmatrix}, &
\varrho_{-\omega_0,\bhv} &\approx\begin{bmatrix}
1 & \bar\epsilon_{\bhv} \\
\bar\epsilon_{\bhv}^* & 0 \end{bmatrix}.
\end{align}
We now switch to the basis of spherical harmonics
$Y_{l,m}(\bhv)$ and define
\begin{align}
\epsilon_{l, m} &= \int Y_{l,m}^*(\bhv)\epsilon_{\bhv}\,
\rmd\Gamma_{\bhv}, &
\bar\epsilon_{l, m} &= \int Y_{l,m}^*(\bhv)\bar\epsilon_{\bhv}\,
\rmd\Gamma_{\bhv}. 
\end{align} 
Using identity
\begin{align}
1-\bhv\cdot\bhv'=4\pi\left[Y_{0,0}(\bhv) Y_{0,0}^*(\bhv')
-\frac{1}{3}\sum_{m=0,\pm1} Y_{1,m}(\bhv) Y_{1,m}^*(\bhv')\right]
\label{eq:1-vv}
\end{align}
one can rewrite Eq.~\eqref{eq:eom-gas} in the linear regime and in the
spherical basis as
\begin{align}
\rmi\partial_t\begin{bmatrix}\epsilon_{l,m} \\
  \bar\epsilon_{l,m}\end{bmatrix}
\approx   \sfLambda_l\cdot
\begin{bmatrix}\epsilon_{l,m} \\
  \bar\epsilon_{l,m}\end{bmatrix},
\end{align}
where
\begin{align}
  \sfLambda_l =
  (1-\alpha)\mu_0\sfI + 
  \begin{bmatrix}
-\eta\omega_0-\mu_l & \alpha\mu_l \\
-\mu_l & \eta\omega_0 + \alpha\mu_l
\end{bmatrix}
\end{align}
with
\begin{align}
\mu_l = \left\{\begin{array}{ll}
\sqrt2\Gf n_\nu & \text{ if } l = 0, \\
-(\sqrt2/3)\Gf n_\nu & \text{ if } l = 1, \\
0 &\text{ otherwise}.
\end{array}\right.
\end{align}
We note that the coupling coefficients
$\mu_l$ of the monopole mode ($l=0$) and the 
dipole modes ($l=1$) have opposite signs.
This sign difference is derived from the current-current nature of the weak
coupling which is proportional to $(1-\bhv\cdot\bhv')$ [see
  Eq.~\eqref{eq:1-vv}]. 

\begin{figure*}[ht]
  \begin{center}
    $\begin{array}{cc}
      \includegraphics*[scale=0.25]{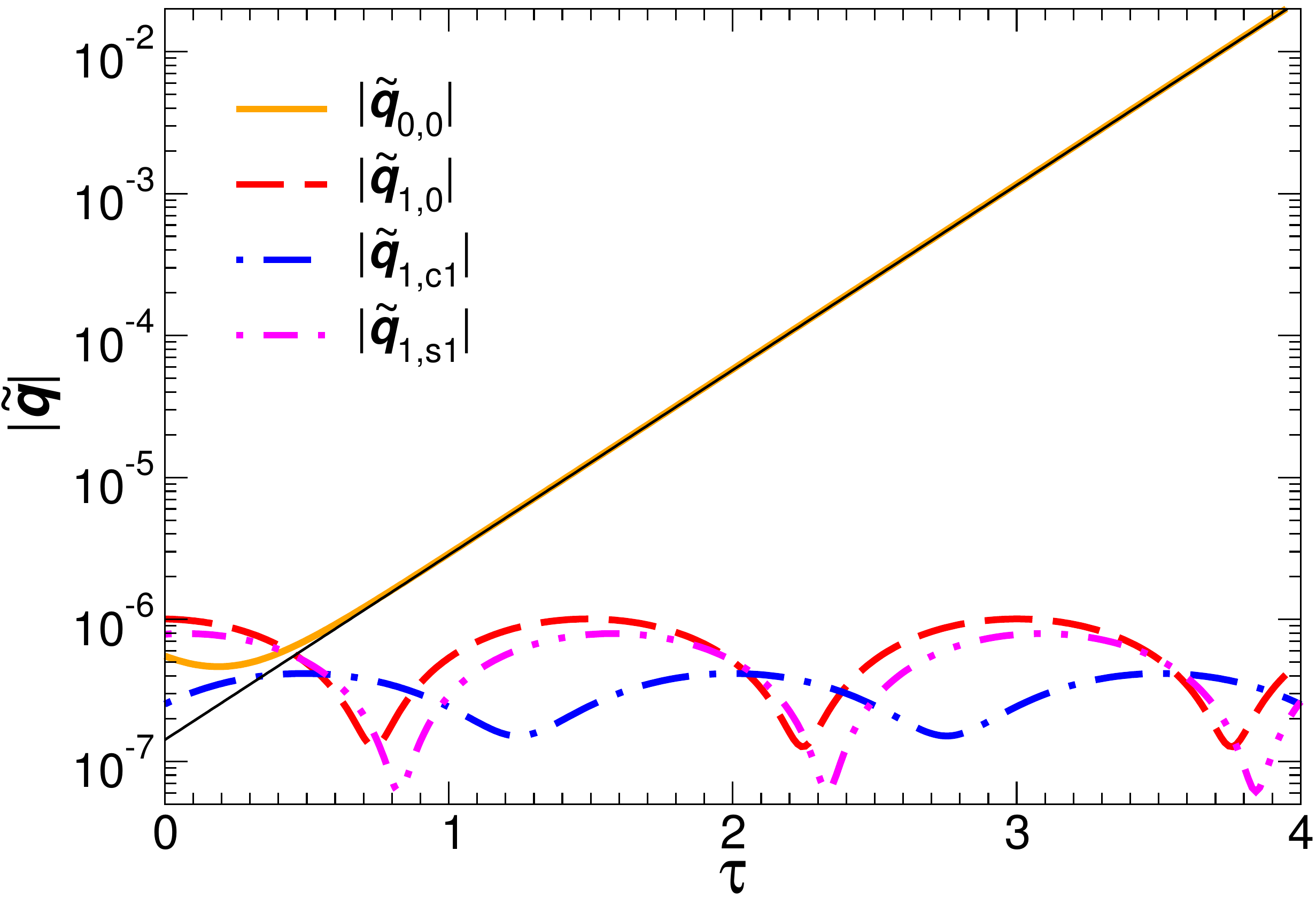} &
      \includegraphics*[scale=0.25]{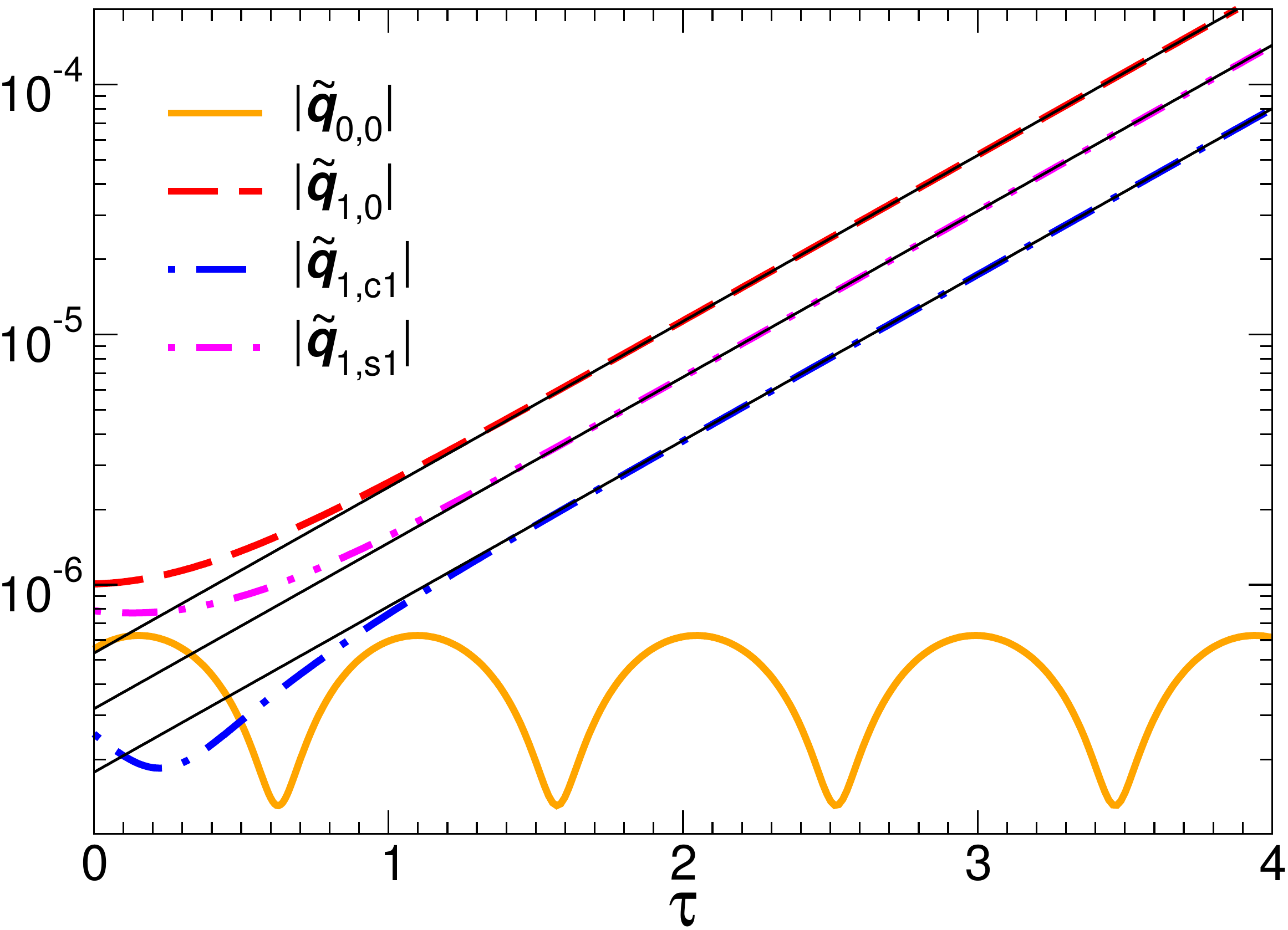}
      \end{array}$
  \end{center}
  \caption{The exponential growth of the amplitude of the monopole
    mode ($|\tilde{\mathbf{q}}_{0,0}|\propto|\epsilon_{0,0}|$) in the
    inverted neutrino mass hierarchy (left)
    and the dipole modes
    ($|\tilde{\mathbf{q}}_{1,0}|\propto|\epsilon_{1,0}|$,
    $|\tilde{\mathbf{q}}_{1,s1}|\propto|(\epsilon_{1,1}-\epsilon_{1,-1})/2\rmi|$ and
    $|\tilde{\mathbf{q}}_{1,c1}|\propto|(\epsilon_{1,1}+\epsilon_{1,-1})/2|$)
    in the normal neutrino mass hierarchy (right) in a homogeneous
    neutrino gas model. The figure is adapted from Fig.~1 of
    Ref.~\citen{Duan:2013kba}.%
    \label{fig:multipoles}}
\end{figure*}

The homogeneous neutrino gas model is 4 dimensional with 1 temporal
dimensions and 3 momentum dimensions. Its e.o.m.\
\eqref{eq:eom-gas} is invariant under a simultaneous but arbitrary
rotation of the momenta of all the neutrinos and antineutrinos. When
this rotation symmetry in momentum space is artificially imposed,
Eq.~\eqref{eq:eom-gas} reduces to the e.o.m.\ of the
2-dimensional bipolar model. The bipolar model permits only the symmetric
solutions which have flavor instabilities only in
IH. However, because $\mu_{l=1}<0$, the dipole modes in the
homogeneous gas model have flavor instabilities in NH [see the
  discussion around Eq.~\eqref{eq:mu-range}] which is clearly shown in
Fig.~\ref{fig:multipoles}. In the regime where the dipole modes become
unstable, the rotation symmetry in the homogeneous gas model is broken
spontaneously which makes it qualitatively different from the bipolar model.

The above result can also be extended to SN models. As mentioned
previously, a
commonly used SN model in collective neutrino oscillations is the 
Bulb model which has the (spatial) spherical symmetry about the
center of the SN and the (directional) axial symmetry about the radial
direction. We define moment basis function 
\begin{align}
\Phi_m(\varphi) = \frac{e^{\rmi\varphi}}{\sqrt{2\pi}}
\qquad(m=0,\pm1,\ldots),
\end{align}
where $\varphi$ is the azimuthal angle of $\bhv$ about the radial
direction.
Using identity
\begin{align}
1-\bhv\cdot\bhv'
=2\pi\left[(1-\cos\vartheta\cos\vartheta')\Phi_0(\varphi)\Phi^*_0(\varphi')
-\frac{1}{2}\sin\vartheta\sin\vartheta'\sum_{m=\pm1}\Phi_m(\varphi)\Phi_m^*(\varphi')\right]
\end{align}
one sees that the coupling coefficients of the $|m|=0$ and $1$ modes
have opposite signs, where $\vartheta$ is the angle that $\bhv$
makes with the radial direction. Similar to the homogeneous neutrino gas mdoel,
only the $m=0$ modes can exist in the Bulb model because of its
imposed axial symmetry. As a result, collective neutrino oscillations
occur only in IH if the neutrino fluxes are bipolar-like.
If the axial symmetry is not imposed, however, collective oscillations
can also occur in NH through the collective modes with $m=\pm1$ which
break the axial symmetry spontaneously. (See
Ref.~\citen{Raffelt:2013rqa} for a detailed analysis and 
Ref.~\citen{Mirizzi:2013rla} for the numerical confirmation of this analysis.)

\subsection{Breaking of the spatial symmetry}
Spatial symmetries can also be broken by collective neutrino
oscillations. This can be illustrated by using the (neutrino) Line
model\cite{Duan:2014gfa}
which is a generalization of the two-beam model in
Sec.~\ref{sec:two-beam}. Compared to the two-beam model, the Line
model does not possess the
translation symmetry along the $x$ axis. But for computational purpose
we will impose a periodic condition
\begin{align}
  \varrho_{\omega,\zeta}(x,z) = \varrho_{\omega,\zeta}(x+L, z),
\end{align}
where $L$ is constant. We will again assume that neutrinos and
antineutrinos are in the (almost pure) electron flavor [see
  Eqs.~\eqref{eq:rho0-two-beam} and \eqref{eq:eps-pm}].
It is convenient to work in the Fourier basis
and define
\begin{align}
  \epsilon_m^\pm(z) &= \frac{1}{L}\int_0^L e^{-\rmi m k_0 x}
  \epsilon^\pm(x,z)\,\rmd x, &
  \bar\epsilon_m^\pm(z) &= \frac{1}{L}\int_0^L e^{-\rmi m k_0 x}
  \bar\epsilon^\pm(x,z)\,\rmd x,  
\end{align}
where $k_0=2\pi/L$. In the linear regime the Fourier modes with
different $m$ values are decoupled, and the linearized e.o.m.\ are 
\begin{align}
\rmi \partial_z\begin{bmatrix}
\epsilon^+_m \\ \bar\epsilon^+_m \\ \epsilon^-_m \\ \bar\epsilon^-_m
\end{bmatrix}
\approx
\sfLambda_m \cdot\begin{bmatrix}
\epsilon^+_m \\ \bar\epsilon^+_m \\ \epsilon^-_m \\ \bar\epsilon^-_m
\end{bmatrix},
\label{eq:line-lin}
\end{align}
where
\begin{align}
  \sfLambda_m=
  v_z^{-1}\begin{bmatrix}
    -\eta\omega_0 -\alpha\mu & \alpha\mu & m q &  0 \\
    -\mu & \eta\omega_0 +\mu & 0 & m q \\
    m q & 0 & -\eta\omega_0 +(2-\alpha)\mu & -\alpha\mu \\
    0 & m q & \mu & \eta\omega_0 + (1-2\alpha)\mu
  \end{bmatrix}
\end{align}
with $q = (2\pi/L)\sqrt{1-v_z^2}$.

The Line model is a 4-dimensional model with 2 spatial dimensions
$(x,z)$ and 2 momentum dimensions $(\omega,\zeta)$. Its e.o.m.\ is
invariant under the translation symmetry along the $x$ 
direction. When this symmetry is artificially imposed, the Line model
reduces to the 3-dimensional two-beam model where only the homogeneous
modes (with $m=0$) are allowed. The Line model allows both the
homogeneous and inhomogeneous ($m\neq0$) modes, and the latter break
the translation symmetry spontaneously when they become unstable.

\begin{figure*}[ht]
  \begin{center}
      \includegraphics*[width=0.7\textwidth]{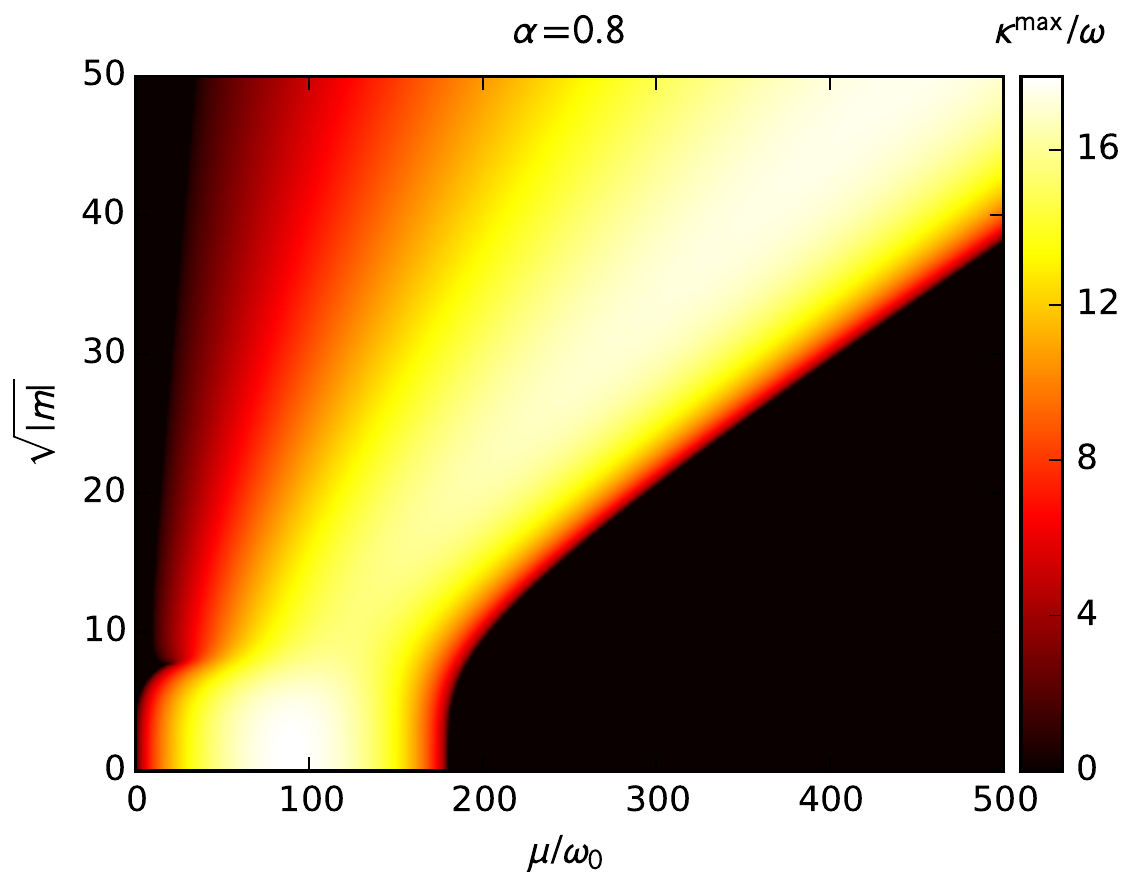}
  \end{center}
  \caption{The flavor instabilities of the Line model with $\alpha =
    0.8$. The color scale represents 
    $\kappa_m^\text{max}(\mu)$, the largest exponential growth rate of
    the collective oscillation modes. The size of the
    periodic box along the $x$-axis is taken to be $L=20\pi/\omega_0$, and the
    propagation directions of the neutrinos are given by unit vectors
    $[v_x,v_z] = [\pm\sqrt3/2,1/2]$. The results are independent of
    the neutrino mass hierarchy. The figure is adapted from Fig.~1 of
    Ref.~\citen{Duan:2014gfa}.%
    \label{fig:kappa}}
\end{figure*}

For given values of $\mu$ and $\alpha$ one can find out if there is
a flavor instability by computing the eigenvalues
of $\sfLambda$. It turns out
that the (two-beam) Line model has the same flavor unstable regions
[in terms of $(\mu,m)$] for both IH and NH.
Fig.~\ref{fig:kappa} shows the flavor unstable region in the $(\mu,m)$
space for the case with $\alpha=0.8$ and $L=20\pi/\omega_0$.
Note that the inhomogeneous modes with large $|m|$ values occur on small
distance scales, and they can be unstable
at larger neutrino densities than the homogeneous mode does.

\section{Summary and outlook}

Neutrino flavor transport in hot and dense astrophysical
environments is a difficult seven-dimensional nonlinear
problem. Using the simplest neutrino gas model, i.e.\ the bipolar
model, we reviewed some of the 
key concepts of collective neutrino oscillations. By applying the
method of linear stability analysis we illustrated that how
the directional and spatial symmetries may be broken spontaneously
during collective neutrino oscillations.

The realization of the spontaneous symmetry breaking commences a new
stage of the study collective neutrino 
oscillations. All the previous research is limited to the models with
imposed symmetries which have smaller dimensions than real
physical systems do. The studies of these simplified models have
yielded important 
insights into the intriguing phenomena of collective neutrino
oscillations, and some of the toy models shall still prove useful in
understanding the results in more realistic models. However, the
incompatibility of the reduced-dimension models and collective
neutrino oscillations clearly points out that these models are
insufficient and that dimensionality really matters in collective
 oscillations. For example, a previous comparison between the results
obtained using the single-angle and
multi-angle SN models shows that neutrino 
oscillations in these two models of different dimensions can have
completely different yields of heavy elements in SN neutrino-driven
wind.\cite{Duan:2010af} The results of the Line model suggest that
neutrino oscillations can occur at larger neutrino densities in the models
of multiple spatial dimensions than in the corresponding models with
only one spatial dimension.
Although a very challenging task, new and realistic multi-dimensional
neutrino gas models must be constructed and employed before we can
know with any confidence what true physical impact neutrino
oscillations may have on astrophysical environments with dense
neutrino media such SNe and black-hole
accretion discs.

\section*{Acknowledgements}
This work was supported by DOE EPSCoR grant DE-SC0008142 at UNM. We thank
S.~Abbar and S.~Shalgar for useful discussions. We also appreciate the
hospitality of INT/UW where part of this work was finished.

\bibliographystyle{ws-ijmpe}
\bibliography{symmetrybreaking.bib}
\end{document}